\documentstyle[aps,epsf,multicol]{revtex}
 
\begin{document}
\author{A. Gammal$^{(a)}$,
T. Frederico$^{(b)}$ and Lauro Tomio$^{(a)}$}
\address{$^{(a)}$
Instituto de F\'\i sica Te\'orica, Universidade Estadual Paulista, \\
01405-900 S\~{a}o Paulo, Brazil \\
$^{(b)}$Departamento de F\'{\i}sica, Instituto Tecnol\'ogico da
Aeron\'autica, \\
Centro T\'ecnico Aeroespacial, \\
12228-900 S\~ao Jos\'e dos Campos,
SP, Brazil }

\title{
Critical number of atoms for attractive Bose-Einstein 
condensates with cylindrically symmetrical traps}
\maketitle

\begin{abstract} 
We calculated, 
within the Gross-Pitaevskii formalism, the 
critical number of atoms for 
Bose-Einstein condensates with two-body attractive interactions in 
cylindrical traps with different frequency ratios. In particular, by using 
the trap geometries considered by the JILA group [
Phys. Rev. Lett. 86, 4211 (2001)],
we show that the theoretical maximum critical numbers 
are given approximately by $N_c = 0.55 ({l_0}/{|a|})$.
Our results also show that, by exchanging the frequencies 
$\omega_z$ and $\omega_\rho$, the geometry with $\omega_\rho < \omega_z$ 
favors the condensation of larger number of particles.
We also simulate the time evolution of the condensate when changing the 
ground state from $a=0$ to $a<0$ using a 200ms ramp. 
A conjecture on higher order nonlinear effects is also added in our
analysis with an experimental proposal to determine its signal and
strength.
\newline\newline  
{PACS: 
03.75.Fi, 32.80.Pj, 11.10.Lm, 02.60.Lj}
\end{abstract}

\begin{multicols}{2}
Bose-Einstein condensates with attractive interactions have been realized
with $^7$Li since 1995 by the Rice group \cite{hulet1} culminating with
experiments that have direct observation of the grow and collapse of this
condensate \cite{huletnature}. Measurements of the maximum critical number 
of atoms $N_c$ in the condensate, in a trap almost spherical, 
were in good agreement with the theoretical predicted numbers,
within the experimental uncertainties.

Recently it has been achieved Bose-Einstein condensation with $^{85}$Rb
\cite{JILA} by means of Feshbach resonance, which allowed wide tunning 
of the scattering length, $a$, from negative to positive. The ability to 
control the scattering length is used to control and measure the stability 
condition with the corresponding critical number of atoms. 

In Ref.~\cite{rup95} it was first shown numerically that for attractive
interactions (negative scattering length $a$) the system becomes unstable 
if a maximum critical number of atoms, $N_c$, is achieved. This 
limit can be stated in a convenient expression by
\begin{equation}
\frac{N_c|a|}{
\sqrt{({\hbar}/{m\omega})}} =k,
\label{k}
\end{equation}
where $m$ is the mass of the particle confined in a trap with 
frequency $\omega$. $k$ is a dimensionless constant, 
directly associated with the critical number of atoms $N_c$.
So, by using the above assumption of a {\it spherically symmetrical 
trap}, several authors \cite{several,huepe}, including us \cite{gammal}, 
have calculated $k$ with a variety of methods. With 
the precision given in Ref.~\cite{gammal}, $k=0.5746$. 
In Ref.~\cite{dalfovo}, it was calculated the critical number
for a nonsymmetrical geometry, but in a case that the frequency ratio
is not too far from the unity ($\omega_z/\omega_\rho = 0.72$), giving
a result for the number of atoms almost equal to the spherical one. 

One can also infer from the variational treatment used in 
Ref.~\cite{huletapp} that the constant $k$ depends on the symmetry of the 
trap. Variational estimates were also considered in Ref.~\cite{wadati}. 
So, in cases of nonspherical symmetry,
the number $k$ will be dependent on the ratios of the trap 
frequencies, with the equation being scaled by some averaged frequency.
As in mostly of the cases considered experimentally
the spatial symmetry is almost cylindrical, with the trap frequencies
given by $\omega_x \approx \omega_y$ and $\omega_z$, we assume
$\omega_\rho = \omega_x = \omega_y $ and a geometrical averaged
frequency given by $\bar\omega = (\omega_z\omega_\rho^2)^{1/3}$.

We define  
\begin{equation}
\lambda\equiv \frac{\omega_z}{\omega_\rho},
\label{aspects}
\end{equation}
such that the trap will have a ``pancake-aspect" if $\lambda>1$;
and a ``cigar-aspect" if $\lambda<1$. The spherical symmetry is
recovered with $\lambda=1$.
It is convenient to redefine the number $k$ given in Eq.~(\ref{k}), 
showing explicitly its dependence on $\lambda$. 
In this case, the critical number of atoms $N_c$ is given by
\begin{equation}
N_c(\omega_\rho,\omega_z) = \frac{k(\lambda)}{|a|}{l_0}
      = \lambda^{-{1}/{6}}\frac{k(\lambda)}{|a|}{l_\rho}
      = \lambda^{{1}/{3}}\frac{k(\lambda)}{|a|}{l_z},
\label{kns}\end{equation}
where $l_0\equiv \sqrt{\frac{\hbar}{m\bar\omega}}$,
$l_\rho\equiv \sqrt{\frac{\hbar}{m\omega_\rho}}$ and
$l_z\equiv \sqrt{\frac{\hbar}{m\omega_z}}$.

Here, in Eq.~(\ref{kns}), we observe explicitly the dependence of
$N_c$ in relation to $\lambda$. 
By exchanging the frequencies $\omega_\rho$ and $\omega_z$ in the trap, we 
observe that $l_\rho \to l_z$, $l_z \to l_\rho$ and $\lambda \to 
1/\lambda$. 
The exchange ratio in this case is given by
\begin{equation}
R(\lambda)\equiv
\frac{N_c(\omega_\rho,\omega_z)}{N_c(\omega_z,\omega_\rho)}
      = \lambda^{{1}/{6}}\frac{k(\lambda)}{k(1/\lambda)}
.\label{R}\end{equation}
$R(\lambda)$ is the relevant factor that affects the number of
particles in the condensate, when exchanging the frequencies
in a cylindrical configuration. In case that $k(\lambda)\sim 
k(1/\lambda)$, we can conclude that $\omega_z >\omega_\rho$ 
results in a larger number of particles inside the trap in the critical 
limit.

The above considerations and the numerical calculations of $k(\lambda)$
that we are communicating are relevant to be taken into account in
experiments with BEC in cylindrical traps with negative $a$, like the
experiments that have been performed in JILA with $^{85}$Rb. 
The JILA's group have considered a ``cigar-type" symmetry 
in their approach~\cite{JILA,JILApp}. 
They have determined, recently, that
$k=0.459\pm0.012$ (statistical) $\pm 0.054$ (systematic), 
for a nonspherical trap, where the frequencies were $17.24\times 
17.47\times 6.80$ Hz. Using the above notation, we can take $\omega_\rho =
\sqrt{\omega_x\omega_y}$ $= 2\pi\times 17.35$ Hz. So, the corresponding
value of $\lambda$ used in Ref.~\cite{JILApp} was 
${\omega_z}/{\omega_\rho} =$ 6.80/17.35 = 0.3919.  

Since the JILA trap is nonspherical, it is worthwhile to determine
numerically the values of $k$, for different $\lambda$.  Our
main goal in the present paper is to systematically calculate $k(\lambda)$
in cylindrical symmetry, either in ``pancake" $(\lambda>1)$ or ``cigar"
type $(\lambda<1)$, in order to verify the favorable geometry of the trap
to condensate a larger number of atoms, when the two-body scattering
length is negative. As we are going to show, the slight discrepancy found
by the JILA group, when comparing their experimental value of $k$ with the
theoretical results, can partially be explained by the present study. 

For an atomic system with negative scattering length and trapped by 
an external harmonic oscillator (non symmetric, in general), 
the Bose-Einstein condensate can be described by 
the Gross-Pitaevskii equation:
\begin{eqnarray}
{\rm i}\hbar\frac{\partial}{\partial t}\Psi(\vec{r},t) &=&
\left[ -\frac{\hbar ^{2}}{2m}{\bf \nabla }^{2}+
\frac{m}{2}\left(
\omega_x^{2}x^{2}+
\omega_y^{2}y^{2}+
\omega_z^{2}z^{2}
\right)\right.\nonumber\\ 
&-& \left.
\frac{4\pi \hbar ^{2}\ |a|}{m}|\Psi (\vec{r},t)|^{2}\right] 
\Psi(\vec{r},t). 
\label{sch}
\end{eqnarray}
The conditions for the validity of this formalism to describe
atomic systems with negative scattering lengths are given in
Ref.~\cite{dalfovo2}.
Deviations due to quantum fluctuations and tunneling, that 
occur near the collapsing region, were studied in
Refs.~\cite{huepe,stoof}. As it appears from such studies, 
the decay probability due to quantum tunneling (that will 
effectively reduce $N_c$) is negligible,
unless $N\approx N_c$.

The wave-function, given by
\begin{equation}
\Psi(\vec{r},t) = \exp{(-{\rm i}\mu
t/\hbar)}
\Psi(\vec{r},0)
\label{psi} ,
\end{equation}
where $\mu$
is the chemical potential,
is normalized to the number of atoms:
\begin{equation}
\int d^3 r |\Psi(\vec{r},t)|^2 = N .
\label{norm}\end{equation}

Using cylindrical symmetry ($\omega_x=\omega_y=\omega_\rho$)
and considering dimensionless units 
[$\tau \equiv \bar\omega t$, $\rho^2 \equiv 
(2m\bar\omega/\hbar)(x^2+y^2)$,
$\zeta ^2 =(2m\bar\omega/\hbar)z^2$], followed by a 
new scaling of the wave-function,
\begin{equation}
\Phi\equiv\Phi(\rho,\zeta;\tau)\equiv 
\sqrt{\frac{4\pi \hbar\ |a|}{m\bar\omega}}
\Psi(\vec{r},t) ,
\label{phi}\end{equation}
we have 
\begin{eqnarray}
{\rm i}\frac{\partial \Phi}{\partial\tau} &=&
\left[ -{\bf \nabla }^{2}+
\left(\frac{\omega_\rho}{\bar\omega}\right)^{2}
\frac{\rho^{2}}{4} +
\left(\frac{\omega_z}{\bar\omega}\right)^{2}
\frac{\zeta^{2}}{4} -
|\Phi|^{2}\right] 
\Phi,
\label{schad}\\
{\rm where}&&\;\;\;\nabla^2\equiv\frac{1}{\rho}
\frac{\partial}{\partial\rho}
\left(\rho\frac{\partial}{\partial\rho}\right)+ 
\frac{\partial^2}{\partial \zeta^2} .\nonumber
\end{eqnarray}

Given the Eqs. (\ref{norm}) and (\ref{phi}), we obtain the 
normalization of $\Phi$ to a defined reduced number of 
atoms $n$:
\begin{equation}
\int_{-\infty}^\infty d\zeta \int_0^\infty 
d\rho\rho |\Phi|^2 = 4\sqrt{2} \frac{N |a|}{l_0} 
\equiv 2 n
\label{normphi},\end{equation}
where, in the critical limit, 
$n=n_c=2\sqrt{2} k$. Eq.(\ref{schad}) depends only on the 
ratio $\lambda=(\omega_z/\omega_\rho)$:
\begin{eqnarray}
\beta \Phi &=&
\left[ -{\bf \nabla }^{2}+
\lambda^{-\frac{2}{3}}
\frac{\rho^{2}}{4} +
\lambda^{\frac{4}{3}}
\frac{\zeta^{2}}{4} -
|\Phi|^{2}\right] \Phi ,
\label{schad2}
\end{eqnarray}
where $\beta\equiv{\mu}/{(\hbar\bar\omega)}$. 
So, the normalization constant $n$, given by (\ref{normphi}), 
as well as $k$, will depend only on $\lambda$.

In our calculation of Eq.~(\ref{schad}) we employed the relaxation method 
propagating in the imaginary time and renormalizing $\Phi$ to $2n$ at 
every step \cite{dalfovo,HJCC}. We searched for stable solutions by 
varying the number $n$ till a critical limit $n_c$. No ground-state 
solutions are possible for $n>n_c$. 
In Fig. 1, we have the corresponding results for the chemical potential as 
a function of $N|a|/l_0 = n/(2\sqrt 2)$. 

\vskip -0.5cm
 \begin{figure}
 \setlength{\epsfxsize}{0.9\hsize}
\centerline{\epsfbox{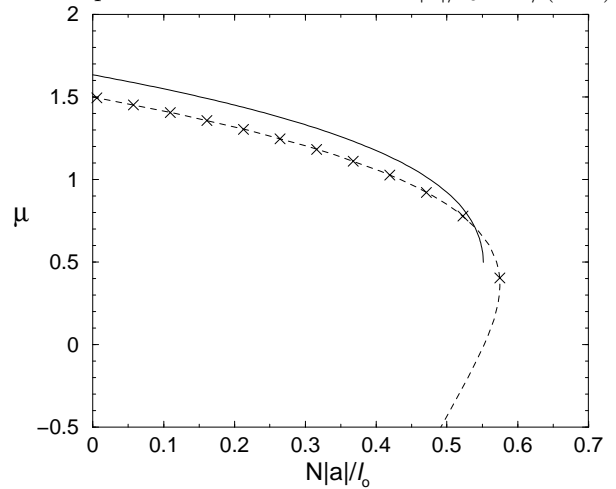}}
 \caption[dummy0]{ 
The chemical potential, $\mu$, is given in units of
$\hbar\bar\omega$, as a function of $N|a|/l_0$. 
Results with spherical symmetry ($\lambda=1$), in dashed line 
and with $\times$, are compared with results using 
$\lambda=6.80/17.35$ (solid line).
Dashed line was obtained using shooting-Runge-Kutta
method, while the $\times$ and the solid line were
obtained by propagation in imaginary time. 
}
\end{figure}

To obtain the results shown in Fig. 1, we first tested our code by running 
the symmetrical case $\lambda=1$ ($\omega_z=\omega_{\rho}$) and comparing 
the results with very precise ones that we have previously calculated 
with the shooting-Runge-Kutta algorithm~\cite{gammal}. 
The plot with $\times$ marks corresponds to the imaginary time
propagation method while the dashed line plot refers to the 
shooting-Runge-Kutta method (in both it was used spherical symmetry). 
One should note that the unstable solutions (back bending branch)
are not accessible by the time-dependent method. 
The plot with solid line shows our results for a cylindrical symmetry,
with the JILA parameters given in \cite{JILApp}, i.e., 
$\omega_z=2\pi\times 6.80$Hz $\omega_\rho=2\pi\times 17.35$ Hz. 
In this case, $k=0.552$ is approximately 4\% lower than the spherical case. 

For the propagation, we have used the Peaceman-Rachford 
alternating-direction implicit method~\cite{PR}.
The time evolution for cylindrical symmetry was performed with a code used 
in \cite{pattanayak}. Our discretization was up to 200 $\times$ 200 in 
$\rho$ and $\zeta$ space directions, and up to 50 in the variable  $\tau$ 
($=\bar\omega t$), with steps of 0.001. 
We also considered $\rho_{max}$ and $\zeta_{max}$ ranging from 2 to 10 
depending on the symmetry. 
In the extreme nonsymmetric cases ($\lambda$ or $1/\lambda$ $>>1$), the
results are more sensible to the grid spacing and to these 
maximum values. In these cases, a lack of precision can occur in the third 
decimal digit of the results shown in Table I.

In Table I, we present the numerical results for the critical constant $k$ 
as a function of the parameter $\lambda=\omega_z/\omega_\rho$, that can 
be useful to analyze experiments with different cylindrical shapes.
Clearly, the optimal value for $k$ occurs for spherically symmetric traps 
($\lambda=1$), as one could also infer from the variational calculations 
given in \cite{huletapp}.
In particular, we determined the values of $k$ for the ratios considered 
in the JILA experiment~\cite{JILA,JILApp}:
The theoretical constant, $k\approx 0.55$, is about 4\% lower than the 
corresponding number with spherical symmetry ($k_s=0.5746$). This can 
partially explain the small disagreement observed in Ref.~\cite{JILApp}
when comparing their result with theoretical ones.

By exchanging the frequencies $\omega_\rho$ and $\omega_z$ in a 
cylindrical symmetry, it is also shown that the ``pancake-type" symmetry
($\omega_z>\omega_\rho$) is preferable (in order to obtain a larger
$N_c$) when $k(\lambda)\approx k(1/\lambda)$.
Considering the exchange ratio presented in Eq.~(\ref{R}) and the results 
shown in Table I for $k(\lambda)$, one can verify the optimal geometry 
to increase the critical number of atoms $N_c$ trapped in a condensate.
Analyzing the ``pancake-type symmetry'', related with the ``cigar-type 
symmetry'' considered by the JILA group in Ref. \cite{JILApp}, 
we note that $\lambda_1 = 17.35/6.80 = 2.5517$, and 
$\lambda_2 = 1/\lambda_1 = 0.3919$. 
As shown in Table I, both cases will give us practically the same constant 
number $k\approx 0.55$.
So, the relevant factor that will decide the convenient symmetry to 
condensate a larger number of atoms is given by Eq.~(\ref{R}), in this 
case; and this favors the ``pancake-type'' geometry:
\begin{equation} 
R(\lambda_1 = 17.35/6.80) \approx 1.17\; .
\end{equation} 
The number of atoms in the condensate can be increased by a
factor of $\sim 17\% $, just by exchanging the geometry of the trap.
The above factor can be verified experimentally, as well as 
other frequency ratios,
with the help of Table I and
the present relations given for $k(\lambda)$ and $R(\lambda)$.        

We should add that other part of the observed discrepancy in the
experimental value of $k$ could be explained
by an early collapse of the condensate due to a dynamical chirp in the 
wave-function when moving the system from $a>0$ to $a<0$. It means
that, when changing the scattering length from a positive to a negative 
value, the energy minimum with $a>0$ is greater than the 
corresponding energy minimum with $a<0$, such that the system 
will collapse at a lower critical number~\cite{chirp}.

We simulated the realistic situation with the parameters given in Ref. 
\cite{JILApp}. We depart from the ground state with $a=0$ and then ramp it 
to $a<0$ in 200ms. In Fig. 2(a) we show the time evolution of the 
mean-square radius, $\rho$, for different final negative scattering 
lengths. For a final value of $N|a|/l_0$ lower or equal to $0.94 k_s$, the 
system presents collective excitations; for a larger value, the system 
collapses. So, we conclude that the dynamical effects can only account 
for about 2\% of the discrepancy observed between the experimental and
theoretical values of $k$. This result implies that, the total correction 
due to the nonspherical symmetrical trap and due to dynamical effects
can only account for a diminishing of about 6\% in the
spherical predicted value of $k$.
For comparison, we also present in Fig. 2(b) the corresponding 
instantaneous shift from $a=0$ to $a<0$. 
\begin{table}
\caption{Numerical solutions for the critical constant $k$, as a 
function of $\lambda=\omega_z/\omega_\rho$. 
$k=k_s$ is for spherical symmetry. With ($\star$) we indicate the symmetry 
considered by the JILA group; alternatively,
with ($\dagger$), the corresponding ``pancake-type'' symmetry.
}  
\begin{tabular}{ccc}
$\lambda $ & $k$ & $k/k_s$ \\
\hline\hline
     0.01     &       0.314   &      0.547 \\          
     0.02     &       0.352   &      0.613 \\         
     0.05     &       0.411   &      0.716 \\          
     0.1      &       0.460   &      0.801 \\        
     0.2      &       0.509   &      0.886 \\         
     0.3      &       0.535   &      0.931 \\        
(6.80/17.35)($\star$) &       0.550   &      0.957 \\        
     0.5      &       0.560   &      0.975 \\        
     2/3      &       0.570   &      0.992 \\        
     1.0      &       0.5746  &      1.000 \\       
     1.5      &       0.570   &      0.992 \\        
     2.0      &       0.561   &      0.976 \\        
(17.35/6.80)($\dagger$)&  0.549   &      0.956 \\       
$\sqrt{8}$    &       0.544   &      0.946 \\        
     3.0      &       0.541   &      0.941 \\        
     4.0      &       0.518   &      0.902 \\       
     5.0      &       0.498   &      0.867 \\       
    10.0      &       0.441   &      0.767 \\       
    20.0      &       0.376   &      0.655 \\       
    50.0      &       0.294   &      0.511 
\end{tabular}
\end{table}
A larger deviation of $k$ is expected in this case, as this numerical 
simulation [shown in Fig. 2(b)] corresponds to a larger chirp in the wave 
function than in the case that $a$ is ``ramping" slowly in time. 
We found that, at $N|a|/l_0 = 0.9 k_s$ the system have complex higher mode 
nonlinear oscillations; for a larger value of $N|a|/l_0$, it collapses. 
So, even in this case, we can account to a maximum of 10\% shift in the 
value of $k$ (including dynamical and nonspherical effects), when 
comparing with the spherical result.

\vskip -.3cm
 \begin{figure}
\setlength{\epsfxsize}{0.9\hsize}
\centerline{\epsfbox{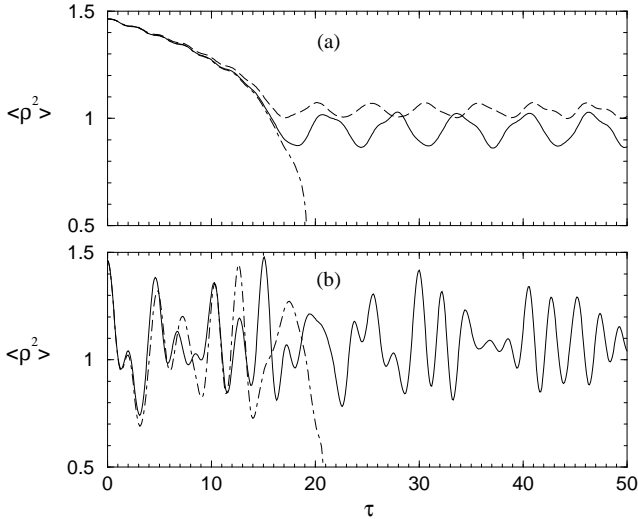}}
\caption[dummy0]{ 
Time evolution of the dimensionless mean-square radius ($\rho$) of the
condensate when changing the ground state from $a=0$ to $a<0$. 
We have considered a 200ms ($\tau=\bar\omega t=16$) linear ramp in 
(a); and an instantaneously shift in (b).
In (a), the dashed, solid and dot-dashed lines correspond to the
ramping until $N|a|/l_0=$ $0.9k_s$, $0.94k_s$, $0.95k_s$, respectively.  
In (b), the dashed line corresponds to the ramping until $N|a|/l_0=$ 
$0.9k_s$; and the solid line corresponds to the ramping until 
$N|a|/l_0=$ $0.91 k_s$.
$k_s$ is the collapse constant $k=N|a|/l_0$ in spherical symmetry. 
Trap parameters were $\omega_{\rho}=2\pi \times 17.35$ Hz and 
$\omega_{z}=2\pi \times 6.80$ Hz. 
}
\end{figure} 
As temperature dependence is being ruled out in the experimental 
analysis, another interesting possibility, that could explain a larger 
deviation in the value of the constant $k$, can be attributed to higher 
order nonlinear effects, that in this case are contributing to increase 
the attractive part of the effective nonlinear potential. 
The relevant effect of a real three-body effective interaction, given by a 
quintic term $g_3|\Phi|^{4}\Phi$ in the rhs of Eq.(\ref{schad2}), 
was already pointed out in \cite{3b}.
If $g_3$ is positive, there is a possibility of 
two-phases in the condensate \cite{3b}.
However, in case that $g_3$ has the same negative sign as the two-body 
interaction, one can also obtain a relevant contribution that may explain 
a smaller value for the constant $k$, as it is occurring in the present 
case. In order to obtain the missing part of deviation ($\sim 10-15$\%), 
we estimated numerically that it is enough to have 
$g_3\approx -0.03$.

A way to obtain some definitive conclusion about the above
conjecture of a relevant role of higher order nonlinearity, is open 
experimentally by 
examining particularly the case $a\approx 0$, when the cubic term in the 
rhs of Eq.(\ref{schad2}) is replaced by a quintic term.
A limit in the number of particles at this particular value of $a$ is a 
good indication of negative higher order nonlinearity; and, given $N_c$, 
the corresponding strength of the nonlinear interaction (that should
mainly come from three-body effects) can be estimated.
\newline
\centerline{\bf Acknowledgments}
\newline
We would like to thank Randall Hulet and Arjendu Pattanayak for useful
discussions.  This work was partially supported by Funda\c{c}\~{a}o de
Amparo \`{a} Pesquisa do Estado de S\~{a}o Paulo and Conselho Nacional de
Desenvolvimento Cient\'{\i}fico e Tecnol\'{o}gico.

\end{multicols} 
\end{document}